\documentclass[a4paper]{jpconf}

\usepackage{graphicx}

\usepackage[numbers,compress]{natbib}

\usepackage{amsmath,amssymb}
\renewcommand\Re{\operatorname{\mathfrak{Re}}}
\renewcommand\Im{\operatorname{\mathfrak{Im}}}

\begin{document}
\title{Studying Deeply Virtual Compton Scattering with Neural Networks%
\footnote{Talk given by K.K. at PHOTON 2011, Spa (Belgium), 22--27 May 2011.}}

\author{Kre\v{s}imir Kumeri\v{c}ki$^1$, Dieter M\"{u}ller$^{2, 3}$, and Andreas Sch\"{a}fer$^4$}

\address{$^1$ Department of Physics, University of Zagreb,
              Bijeni\v{c}ka c. 32, 10002 Zagreb, Croatia}
\address{$^2$ Lawrence Berkeley National Lab, One Cyclotron Rd, Berkeley, CA 94720, U.S.A.}
\address{$^3$ Institut f\"ur Theoretische Physik II, Ruhr-Universit\"{a}t Bochum,
              Universit\"{a}tsstra\ss{}e 150, 44780 Bochum, Germany}
\address{$^4$ Institut f\"{u}r Theoretische Physik, Universit\"{a}t Regensburg,
              Universit\"{a}tsstra\ss{}e 31, 93053 Regensburg, Germany}

\begin{abstract}
Neural networks are utilized to fit Compton form factor $\mathcal{H}$ to HERMES data
on deeply virtual Compton scattering off unpolarized protons. We used this
result to predict the beam charge-spin assymetry for muon scattering off proton at the
kinematics of the COMPASS II experiment.
\end{abstract}

\section{Introduction}

Deeply virtual Compton Scattering (DVCS) is recognized as the theoretically
cleanest process for accessing generalized parton distributions (GPDs)
\cite{Mueller:1998fv,Radyushkin:1996nd,Ji:1996nm}, which
describe the three-dimensional structure of nucleon in terms of partonic degrees of
freedom. Determination of GPDs, beside improving our general understanding
of QCD dynamics, allows to address important questions such as
the partonic decomposition of the nucleon spin  \cite{Ji:1996ek} and
characterization of multiple-hard reactions in
proton-proton collisions at LHC collider \cite{Diehl:2011tt}.
Concerning the latter, GPD-describable non-trivial transversal structure of proton,
such as the correlation between parton's longitudinal momentum fraction and its
transversal distance is already finding its way in popular event generators, such
as PYTHIA \cite{Corke:2011yy}.

Similarly to extraction of normal parton distribution functions (PDFs), extraction
of GPDs can be performed by global model or local fits to available data
\cite{Kumericki:2007sa,Guidal:2008ie,Kumericki:2009uq,Guidal:2009aa,Guidal:2010ig, Moutarde:2009fg}.
However, compared to global PDF fits, extracting GPDs from data is a much more
intricate task and model ambiguities are much larger.
Due to the facts that GPDs cannot be fully constrained from data and that they
depend at the input scale  on three variables, the space
of possible functions, although restricted by GPD constraints, is huge. As a result,
the theoretical systematic error induced by the choice of the fitting model
is much more serious than in the PDF case, where the model functions
depend at the input scale only on one variable, namely, the longitudinal momentum fraction $x$.

Here we report on some results obtained using alternative approach \cite{Kumericki:2011rz}, in which
\emph{neural networks} are used in place of specific models.
This essentially eliminates the problem of model dependence
and, as an additional advantage, facilitates a convenient method to propagate
uncertainties from experimental measurements to the final result.
Our approach is mostly similar to the one already employed
for $F_2$ structure function and PDF extraction
\cite{Forte:2002fg,Ball:2008by,Ball:2010de}
and will be shortly described in the next section.
To reduce the mathematical complexity of the problem we have fitted not the GPDs
itself, but the dominant Compton form factor (CFF) $\mathcal{H}(x_B, t)$, depending on
Bjorken variable $x_B$ and momentum transfer $t$. At leading order,
the imaginary part of this CFF is related to the corresponding GPD $H(x, \xi, t)$
at the cross-over line $x=\xi$:
\begin{equation}
\Im{\cal H}(x_B=\frac{2\xi}{1+\xi},t)
\stackrel{\rm LO}{=}
\pi \bigg( H(\xi,\xi,t) - H(-\xi, \xi,t) \bigg)  \;,
\label{eq:DVCSLO}
\end{equation}
so knowledge of CFF $\mathcal{H}$ provides us with direct information about the
proton structure.

\section{\label{sec:nnmethod} The Method of Fitting Data with Neural Networks}


\begin{figure}[th]
\begin{center}
\includegraphics[scale=0.8]{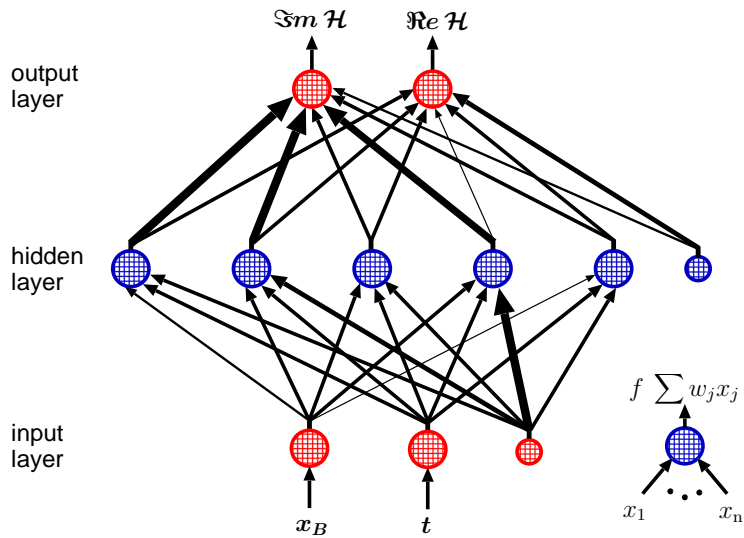}
\end{center}
\caption{A neural network  parametrization of
the complex-valued CFF $\mathcal{H}(x_B, t)$. Each blob symbolizes a neuron
and thickness of arrows represents the strengths of weights $w_j$.}
\label{fig:perceptron}
\end{figure}

The neural network type used in this work, known as \emph{multilayer perceptron},
is a mathematical structure consisting of a number of interconnected
``neurons'' organized in several layers.
It is schematically shown on Figure~\ref{fig:perceptron}, where each blob
symbolizes a single neuron.
Each neuron has several inputs and one output. The value
at the output is given as a function $f(\sum_j w_j x_j)$ of a sum of values at inputs
$x_1, x_2, \cdots$, each weighted by a certain number
$w_j$. For \emph{activation function} $f(x)$ we employed logistic sigmoid function $$f(x) = 1/(1+\exp(-x))$$
for neurons in inner (``hidden'') layer, while for input and output layers we
used the identity function.

By iterating over the following steps the network is trained, \emph{i.e.}, it ``learns'' how to describe
a certain set of data points:
\begin{enumerate}
\item Kinematic values (two in our case: $x_B$ and $t$) of
the first data point are presented to two input-layer neurons
\item These values are then propagated through the network
according to values of weights $w_j$.
In the first iteration, weights are set to some random value.
\item As a result, the network produces some resulting values
of output in its output-layer neurons. Here we have
two: $\Im\mathcal{H}$ and $\Re\mathcal{H}$.
Obviously, after the first iteration, these will be
some random functions of the input kinematic values:
$\Im\mathcal{H}(x_B, t)$ and $\Re\mathcal{H}(x_B, t)$.
\item Using these values of CFF(s), the observable corresponding
to the first data point is  calculated and it is  compared to
actually measured value, with squared error used
for building the standard $\chi^2$ function.
\item The obtained error is then used to modify the network:
It is propagated backwards
through the layers of the network and each weight is adjusted
such that this error is decreased.
\item This procedure is then repeated with the next data point, until
the whole training set is exhausted.
\end{enumerate}
This sequence of steps\footnote{The described procedure, where the neural
network is modified after each addition of a data point is known as 
\emph{sequential learning}. In the alternative procedure, \emph{batch learning}, 
the network is modified only after the
complete training set is presented to the network and the
total error is calculated, \emph{i.e.}, last two steps of the above procedure are
reversed. We tried both types of learning and batch learning turned out to be
more robust.} is repeated
until the network is capable to describe experimental data with a sufficient
accuracy. To guard against overfitting the data (``fitting to the noise''),
one (randomly chosen) subset of data is not used for training but only for monitoring the
progress and stopping the training when error of network description of
this data starts to increase significantly. This ensures that resulting
neural network represents a function which is not too complex and which
is thus expected to provide a reasonable estimate of the actual underlying
physical law.

To propagate experimental uncertainties into the final result we used the
``Monte Carlo'' method \cite{Giele:2001mr} where neural networks are not
trained on actual data but on a collection of ``replica data sets''. These sets
are obtained from original data by generating random artificial data points
according to Gaussian probability distribution with a width defined by the
error bar of experimental measurements.  Taking a large number $N_{rep}$ of
such replicas, the resulting collection of trained neural networks
$\mathcal{H}^{(1)},\ldots,\mathcal{H}^{(N_{rep})}$
defines a probability distribution $\mathcal{P}[\mathcal{H}]$ of the represented CFF
$\mathcal{H}(x_B, t)$ and of any functional $\mathcal{F}[\mathcal{H}]$
thereof. Thus, the mean value of such a functional and
its variance are \cite{Giele:2001mr,Forte:2002fg}
\begin{align}
 \Big\langle \mathcal{F}[\mathcal{H}] \Big\rangle& =
 \int \mathcal{D}\mathcal{H}
 \: \mathcal{P}[\mathcal{H}] \, \mathcal{F}[\mathcal{H}] =
  \frac{1}{N_{rep}}\sum_{k=1}^{N_{rep}} \mathcal{F}[\mathcal{H}^{(k)}]\;,
\label{eq:funcprob} \\
\Big(\Delta \mathcal{F}[\mathcal{H}]\Big)^2& =
\Big\langle \mathcal{F}[\mathcal{H}]^2 \Big\rangle  -
\Big\langle \mathcal{F}[\mathcal{H}] \Big\rangle^2 \;.
\label{eq:variance}
\end{align}
More details about our procedure can be found in \cite{Kumericki:2011rz}.

\section{\label{sec:fit} First results}

We now present neural network fits \cite{Kumericki:2011rz} to two sets of
HERMES collaboration measurements \cite{:2009rj} of leptoproduction
of a real photon by scattering leptons off
unpolarized protons (of which DVCS is a subprocess).
One set consists of 18 measurements of the first sine harmonic
$A_{LU}^{\sin\phi}$ of the
beam spin asymmetry (BSA)
\begin{equation}
BSA
\equiv \frac{ {\rm d}\sigma_{e^\uparrow} - {\rm d}\sigma_{e^\downarrow} }
{ {\rm d}\sigma_{e^\uparrow} + {\rm d}\sigma_{e^\downarrow} }
\sim
A_{LU}^{\sin\phi}\sin\phi \;,
\label{eq:bsa}
\end{equation}
(where $\phi$ is the azimuthal angle in the so-called Trento convention),
while in the other set there are 18 measurements of the first
cosine harmonic $A_{C}^{\cos\phi}$ of the beam charge asymmetry (BCA)
\begin{equation}
BCA
\equiv \frac{ {\rm d}\sigma_{e^+} - {\rm d}\sigma_{e^-} }
{ {\rm d}\sigma_{e^+} + {\rm d}\sigma_{e^-} }
\sim
A_{C}^{\cos 0\phi} + A_{C}^{\cos\phi}\cos\phi \;.
\label{eq:bca}
\end{equation}
Both sets cover the identical kinematic region
$$0.05 < x_B < 0.24\,,\quad
0.02  < -t/{\rm GeV}^2< 0.46\,,\quad \mbox{and}\quad
1.2 < \mathcal{Q}^2/{\rm GeV}^2 < 6.11\,,$$
and in this region BSA and BCA
are determined essentially by the imaginary and the real part of the
Compton form factor $\mathcal{H}$ \cite{Belitsky:2001ns}, respectively,
so we ignored other CFFs.
Furthermore, in this region the dependence of $\mathcal{H}$ on the photon
virtuality $\mathcal{Q}^2$ is weak and,
therefore, we neglected it for simplicity.
Thus, at present, just a single CFF $\mathcal{H}(x_B, t)$, or two real-valued
functions $\Im\mathcal{H}(x_B, t)$ and $\Re\mathcal{H}(x_B, t)$, are extracted from
data by neural networks.

\begin{figure}[t]
\begin{center}
\includegraphics[scale=0.52,clip]{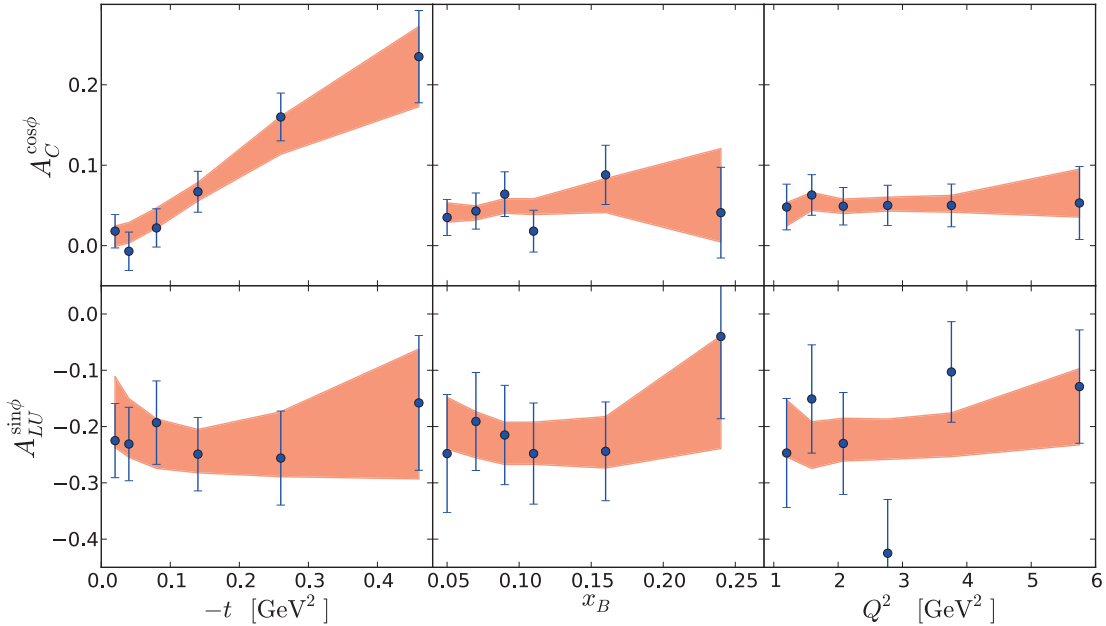}
\end{center}
\caption{First cosine harmonic of beam charge asymmetry $A_{C}^{\cos\phi}$
(\protect\ref{eq:bca}) and first sine
harmonic of beam spin asymmetry $A_{LU}^{\sin\phi}$ (\protect\ref{eq:bsa})
resulting from neural network fit,
shown together with data \cite{:2009rj}, used for training.
\label{fig:hermes09}}
\end{figure}

We fitted this data using the method described in the previous section where we
constructed 50 neural networks with architecture (2-13-2), \emph{i.e.}, with
two input neurons (for two kinematic variables $x_B$ and $t$), 13
neurons in the hidden layer (where we convinced ourselves that adding or
removing few neurons doesn't significantly change the results), and 2 output
neurons (representing $\Im\mathcal{H}(x_B, t)$ and $\Re\mathcal{H}(x_B, t)$).
On Figure~\ref{fig:hermes09} we show the fit quality by presenting
the data used for training together with the description of this data
by the final set of 50 neural networks, using relations
(\ref{eq:funcprob}) and (\ref{eq:variance}).

\begin{figure}[t]
\centerline{\includegraphics[scale=0.5,clip]{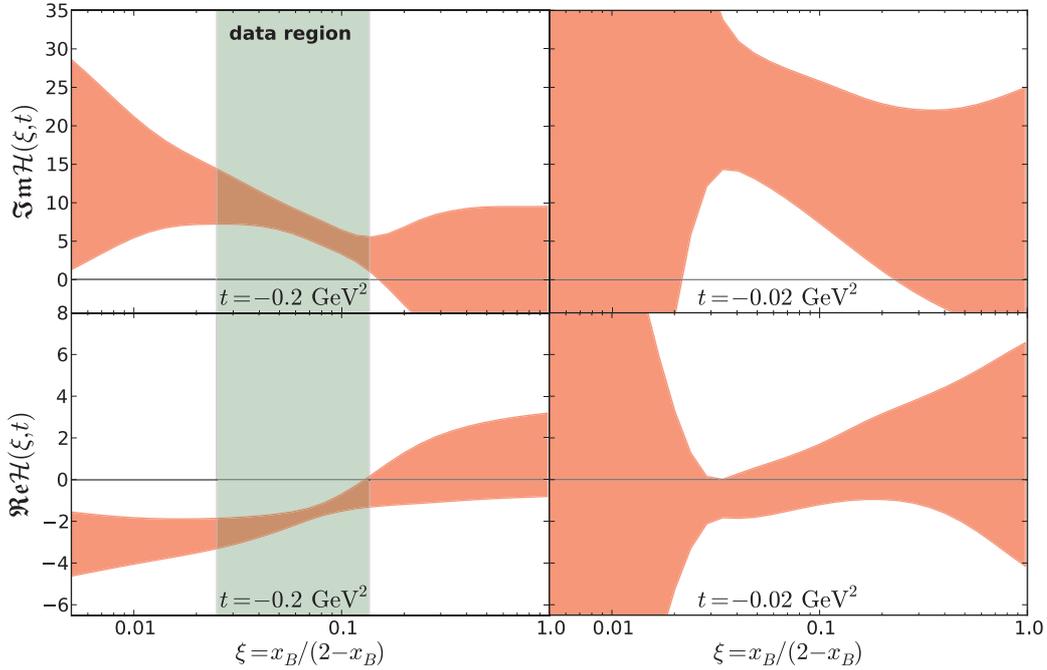}}%
\caption{\small
Neural network extraction of $\Im{\cal H}(x_{\rm Bj},t)$
and $\Re{\cal H}(x_{\rm Bj},t)$
from HERMES BCA and BSA~\cite{:2009rj} data.
Actual data region is shown as vertical band in the
middle of the left two panels. Outside of this band
and on the whole of the right two panels, neural
networks are \emph{extrapolating} from the data.
\label{fig:cff} }
\end{figure}

CFF $\mathcal{H}$ itself, which is our main result, is displayed in
Figure~\ref{fig:cff},  where
$\Im\mathcal{H}$ (upper panels) and $\Re\mathcal{H}$ (lower panels) are separately plotted.
One notices that in the kinematic region of the measured data (this is roughly the
middle vertical third of the left panels), where neural networks are
\emph{interpolating} the data, CFF $\mathcal{H}$ is estimated with
a reasonably small uncertainty.
However, as one starts to \emph{extrapolate} the fitted CFF $\mathcal{H}$ outside of the
data region (left and right thirds of left panels
and the whole of the right panels), the neural network parameterization of
CFF $\mathcal{H}$ is very unconstrained.
This is particularly visible in the right panels, illustrating
the difficulty of a model-independent extrapolation
towards $t= 0$, which is a limit of particular
interest for hadron structure studies.

\begin{figure}[t]
\begin{center}
\includegraphics[scale=0.5,clip]{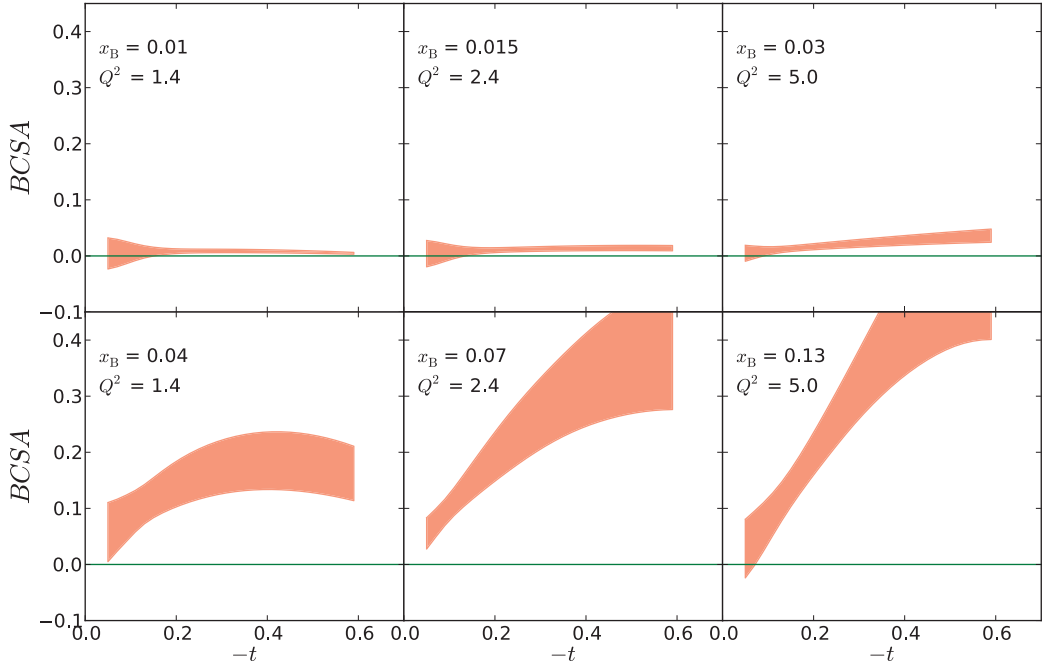}
\end{center}
\caption{Predictions of neural network fits for beam charge-spin asymmetry
(\ref{eq:bcsa}) in COMPASS II kinematics
($E_{\mu} =160\, {\rm GeV}$,  $\phi = 0$).}
\label{fig:compass}
\end{figure}

Finally, as an example of a proper prediction coming from our analysis
we plot in Figure~\ref{fig:compass} the beam charge-spin asymmetry (BCSA),
\begin{equation}
BCSA
\equiv \frac{ {\rm d}\sigma_{\mu^{\downarrow +}} - {\rm d}\sigma_{\mu^{\uparrow -}} }
{ {\rm d}\sigma_{\mu^{\downarrow +}} + {\rm d}\sigma_{\mu^{\uparrow -}} } \;,
\label{eq:bcsa}
\end{equation}
as a function of momentum transfer $t$, for several kinematic points that are
characteristic for the COMPASS II experiment of scattering muons and
antimuons off proton target (where the muon is taken to
be massless and the polarization is set equal to 0.8).
This experiment was chosen because its kinematics overlap with that of the
HERMES data used for neural network training. Hence these predictions
represent partly interpolation and partly extrapolation of HERMES data,
thus testing this whole approach in a nontrivial way.

\section{Conclusions and outlook}

By explicit extraction of Compton form factor $\mathcal{H}$
from HERMES data on beam spin and charge asymmetries
we demonstrated that neural networks can be a powerful tool for
studying hadron structure.
They can interpolate experimental data in an unbiased way,
eliminating thus the systematic error introduced by choosing a
specific fitting function in the standard model-fitting approaches.
Since GPDs and CFFs are multivariate functions, this advantage
is much more pronounced than in PDF fitting, where PDFs
on the input scale depend only on a single variable.
Still, this feature of neural network approach cuts both ways:
unbiased fitting of GPDs and CFFs to the precision of the
present PDF fits would require orders of magnitude more
data then presently available --- to cover the larger
dimension of the kinematic space.
To overcome this problem, one may deliberately introduce some
biases and constraints on neural networks, especially those that correspond
to certain well established properties of represented functions
(e.g. dispersion relations between imaginary and real parts
of CFFs \cite{Teryaev:2005uj,Kumericki:2007sa,Diehl:2007jb,Kumericki:2008di}).
Furthermore, one could view neural network fits as
intermediate results and use them as a tool for
model-dependent studies.

\section*{Acknowledgments}

This work was supported by the BMBF grant under the contract no. 06RY9191,
by EU FP7 grant HadronPhysics2, by DFG grant, contract no. 436 KRO 113/11/0-1 and by
Croatian Ministry of Science, Education and Sport, contract no.
119-0982930-1016.

\providecommand{\newblock}{}

\end{document}